\documentclass[aps,preprint,showpacs,amsmath,amssymb]{revtex4-1}
\usepackage{graphicx}
\usepackage{dcolumn}
\usepackage{bm}
\usepackage{hyperref}
\begin{document}

\title{Genuinely entangling uncorrelated atoms via Jaynes-Cummings interactions}
\author{Mazhar Ali}
\affiliation{Department of Electrical Engineering, Faculty of Engineering, Islamic University Madinah, 107 Madinah, Saudi Arabia}


\begin{abstract}
We study three independent pairs of Jaynes-Cummings systems such that two atoms might be correlated with each other but the third atom 
is uncorrelated with rest. We investigate the conditions under which these uncorrelated three atoms may become genuinely entangled. We 
find that this task is impossible if the cavity interacting with uncorrelated atom share classical correlations with any other cavity. 
We observe that atomic state can become genuine multipartite entangled, at least if the cavity with uncorrelated atom, is highly entangled 
with any other cavity. This is an interesting and non-trivial observation and may serve as another technique to generate multipartite 
entangled atoms via JC-interactions. The findings can be realized with available experimental setups. 
\end{abstract}

\pacs{03.65.Yz, 03.65.Ud, 03.67.Mn}

\maketitle

\section{Introduction}\label{S-intro}

Quantum theory for atom-field interactions predicted certain peculiar possibilities which were later observed in experiments. Initial findings were 
spontaneous emission of excited atoms interacting with vacuum, collapse and revival of atomic population, micro-masers, lasers, etc. 
The simplest model consist of a two-level atom interacting with quantized electromagnetic field with strong coupling constant. This so called 
"toy" model of matter-field interaction is well known in literature as Jaynes-Cummings (JC) model. It is well known that atom-field interaction 
gives a combined atom-field entangled state. This single JC-model was extended to "double-JC" model in which two independent atoms initially entangled 
with each other were allowed to interact only with their own cavities. It was shown that entanglement either oscillated periodically or exhibited 
collapse and revival phenomenon depending upon initial state \cite{Yonac-JPB39-2006}. After this study, several authors have applied this simple model 
to study quantum correlations of atomic systems or of field states or for various partitions of double or triple JC-systems 
\cite{Li-JPB40-2007, Man-JPB41-2008, Qiang-JPB48-2015, Wang-SR7-2017, Qiang-QIP17-2018, Ghoshal-PRA101-2020}. In a recent work \cite{Qiang-QIP17-2018},  
the authors have studied multipartite concurrence for triple JC model. They have considered the three two-level atoms to be in a specific 
genuine multipartite entangled state such that each atom interacts with their own independent cavity. They have shown that concurrence of three atoms 
exhibits collapse and revivals in a periodic fashion. They have also found that three independent cavities can become genuinely entangled as 
entanglement is transferred from the atoms to cavities. More recently, Ding et. al \cite{Ding-arx2020} have taken two identical two-level atoms in a 
common cavity undergoing JC interaction and have shown the freezing and thawing of entanglement in this three qubit system. 

In the current work, we raise and answer a different question which to our knowledge has not been studied so far. We consider the triple JC model in 
which three two-level atoms (as qubits) are independent and uncorrelated with each other in the start of interaction (in a tensor product state 
instead of a genuine entangled state). The two atoms can be correlated with each other but third atoms is not correlated with other atoms. 
Similarly, two cavities can share some correlations but one of cavity is uncorrelated with others. Each of the atom interacts with its own independent 
cavity. Collectively, atomic as well as cavities states are uncorrelated. In particular, we do not have any genuine entanglement in the system at 
the start. We ask the question that whether it is possible to entangle all three atoms such that the resulting atomic state is not bi-separable. 
Surprisingly, we find that three atoms can become genuinely entangled. However, this influence is only possible whenever the cavity containing 
uncorrelated atom must share an entangled state with any of other two cavities. We show that it is impossible for three atoms to become genuinely 
multipartite entangled, if the cavity containing uncorrelated atom share only classical correlations with any other cavity. 
We also explore the possibility of uncorrelated atom to get entangled with either of the other two atoms. We find that two atoms which are 
interacting independently with their cavities (correlated ones) may get entangled, however, we find no evidence of entanglement between the atom in 
uncorrelated cavity with the uncorrelated atom. We also find that although classical correlations shared between two cavities can not support genuine 
entanglement among three atoms, nevertheless, there do exist other kinds of quantum correlations among three atoms, and among any pair of atoms. We 
aim to study such quantum correlations in a separate report. 

This paper is organized as follows. In section \ref{Sec:Model}, we briefly describe our model of interaction, equation of motion and the method 
to obtain the general solution. We review the concept of separability, bi-separability, and genuine entanglement in section \ref{Sec:GME}, where 
we also describe the criteria to detect such correlations in time evolved atomic state. We present our findings in section \ref{Sec:results} and 
provide a summary of our work in section \ref{Sec:conc}. 

\section{Triple Jaynes-Cummings interactions: Model and equation of motion} \label{Sec:Model}

\begin{figure}[h]
\centering
\scalebox{2.35}{\includegraphics[width=1.99in]{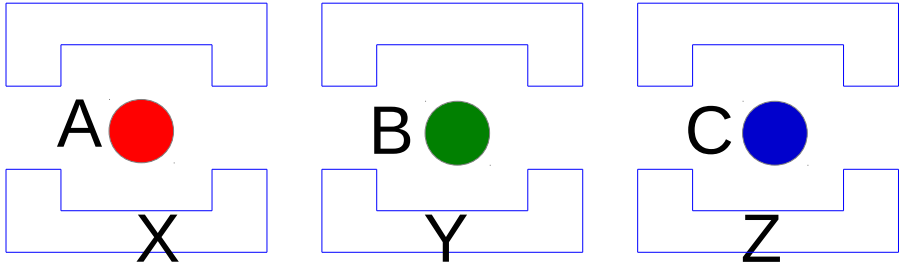}}
\caption{(Color online) Three two level atoms, $A$, $B$, and $C$ are interacting with cavities $X$, $Y$, and $Z$, respectively. There is 
no direct interaction among atoms as well as cavities.}
\label{FIG:model}
\end{figure}

In this section, we illustrate our physical model and the basic equation of motion. We consider three two-level atoms (qubits) $A$, $B$, and $C$, 
such that each atom is interacting with its own cavity $X$, $Y$, and $Z$, respectively. Figure~\ref{FIG:model} shows this triple JC system. We 
assume that neither the atoms nor the cavities have any direct interaction among themselves. This means that each JC system is independent and 
have no direct interaction with other JC systems. We allow initial correlations between atoms $A$ and $B$ as well as between cavities $Y$ 
and $Z$. Atom $C$ (and cavity $X$) is uncorrelated with other atoms (cavities) at the start. We will show below that these correlations are responsible 
for some interesting dynamics in this system. In the absence of such correlations, three JC systems would be completely uncoupled and time evolution of 
such triple JC system would not develop any correlations with other parts of the system.   

We take atoms $A$ and $B$ to be in Werner states, whereas the third atom $C$ is uncorrelated with rest of the system at the start of the interaction. 
The initial combined state of three atoms can be written as 
\begin{eqnarray}
\varrho^{ABC} (0) = \varrho^{AB} \otimes \varrho^C  \, , 
\end{eqnarray}
where the state of atoms $A$ and $B$ is given as  
\begin{eqnarray}
\varrho^{AB} = \alpha \, | \psi^- \rangle\langle \psi^- | + \frac{1 - \alpha}{4} \, \mathbb{I}_4 \, , 
\end{eqnarray}
with $|\psi^-\rangle = 1/\sqrt{2} (|0, 1\rangle - | 1, 0\rangle)$ is one of maximally entangled Bell state, $\mathbb{I}_4$ is identity matrix, 
and $0 \leq \alpha \leq 1$. It is clear that atoms $A$ and $B$ would be sharing only classical correlations for $\alpha = 0$. For 
$ 0 < \alpha \leq 1/3$, the atoms also share some quantum correlations (like quantum discord, or local quantum uncertainty, etc) and in addition, two 
atoms are entangled for $1/3 < \alpha \leq 1$. It is sufficient to take the third atom in state 
\begin{eqnarray}
\varrho^{C} = | \phi \rangle\langle \phi | \, , 
\end{eqnarray}
with $\phi = \beta \, |0\rangle + \sqrt{1 - \beta^2} \, | 1\rangle$, (with $0 \leq \beta \leq 1$) which is a superposition of ground and excited 
state for atom $C$. Similarly, we allow cavities $Y$ and $Z$ to share correlations with each other but cavity $X$ is uncorrelated with any atom or 
cavity at the start of interactions. The combined state of three cavities can be written as
\begin{eqnarray}
\varrho^{XYZ} (0) = \varrho^{X} \otimes \varrho^{YZ} \, , 
\end{eqnarray}
where we have defined the state of cavities $Y$ and $Z$ similar to atomic state $\varrho^{AB}$ as 
\begin{eqnarray}
\varrho^{YZ} = \gamma \, | \psi^- \rangle\langle \psi^- | + \frac{1 - \gamma}{4} \, \mathbb{I}_4 \, , 
\end{eqnarray}
with $0 \leq \gamma \leq 1$, and 
\begin{eqnarray}
\varrho^{X} = | \zeta \rangle\langle \zeta | \, , 
\end{eqnarray}
with $\zeta = \kappa \, |0\rangle + \sqrt{1 - \kappa^2} \, | 1\rangle$ ($0 \leq \kappa \leq 1$), which is a superposition of zero and single 
excitation in cavity $X$. The initial combined state of three atoms and three cavities is given as
\begin{equation}
\varrho_{XYZ}^{ABC} (0) = \varrho^{ABC} \otimes \varrho^{XYZ} \, .
\end{equation}

It is clear from this situation that if cavity $Z$ share no correlations of any kind with other two cavities, then the single JC system 
(atom $C$ + cavity $Z$) would have nothing to do with the rest of atoms and cavities. So the only way for other atoms and cavities to "talk" with 
atom $C$ is via correlations, which cavity $Z$ may share in the system. We are interested in the following questions: 
Is it possible that atomic state $\varrho^{ABC}(t)$ can become genuine multipartite entangled (defined in next section)? If 
yes, then what kind of correlations (classical, discord or entanglement) the cavity $Z$ must share with cavity $Y$ for this phenomenon to happen? 
In addition, we can also ask under what conditions, atom $C$ can be entangled with any of the other two atoms, $A$ or $B$? To answer these questions, 
we must first describe the interaction model of atom and field. 

The Hamiltonian describing this interaction can be written as \cite{Qiang-QIP17-2018}
\begin{eqnarray}
\hat{H} &=& \hbar \sum_{k = A, B, C} \, \bigg[ \nu_k \hat{a_k}^\dagger \hat{a}_k + \frac{\omega_k}{2} \sigma_z^k 
+ g_k (\hat{a}_k^\dagger \sigma_-^k + \hat{a}_k \sigma_+^k ) \,, \bigg]
\label{Eq:Hamil}
\end{eqnarray}
where $\nu_k$ is the field frequency, $\omega_k$ is the energy difference between two atomic levels, $g_k$ is the atom-field coupling 
constant, $\hat{a}_k^\dagger (\hat{a}_k)$ is the creation (annihilation) operator for the cavity field, 
$\sigma_z^k = |1_k\rangle\langle 1_k | - |0_k\rangle\langle 0_k |$, and 
$\sigma_+^k = |1_k\rangle\langle 0_k | (\sigma_-^k = |0_k\rangle\langle 1_k |)$ is the atomic raising (lowering) operator. For greater simplicity
in calculations and notations, we have considered the zero detuning $\Delta = \nu_k - \omega_k = 0$, also called resonance condition. 
In the unitary time evolution method, we can write (for any single atom-field interaction) 
\begin{equation}
\hat{U}_k = e^{- i \, \frac{\mathcal{H}_k \, t}{\hbar} }\, ,
\end{equation}
where $\mathcal{H}_k$ is the Hamiltonian in interaction picture. This unitary operator can be written as \cite{Scully-Zubairy}
\begin{eqnarray}
\hat{U}_k = \cos (g_k t \sqrt{\hat{a}_k^\dagger \hat{a}_k + 1}) |1_k\rangle\langle 1_k | 
+ \cos (g_k t \sqrt{\hat{a}_k^\dagger \hat{a}_k}) |0_k\rangle\langle 0_k | \nonumber \\  
- i \frac{\sin (g_k t \sqrt{\hat{a}_k^\dagger \hat{a}_k + 1})}{\sqrt{\hat{a}_k^\dagger \hat{a}_k + 1}} \hat{a}_k |1_k\rangle\langle 0_k | 
- i \frac{\sin (g_k t \sqrt{\hat{a}_k^\dagger \hat{a}_k + 1})}{\sqrt{\hat{a}_k^\dagger \hat{a}_k + 1}} \hat{a}_k^\dagger |0_k\rangle\langle 1_k | \,. 
\label{Eq:SWE}
\end{eqnarray}
The time evolution of an arbitrary initial atom-field state evolves as $ \varrho (t) = \hat{U}_k \, \varrho (0) \hat{U}_k^\dagger$,   
where $ \varrho (0)$ is the initial state of a two-level atom and a cavity. In our case, the combined state of three atoms and their 
respective cavities evolve as 
\begin{eqnarray}
\varrho_{XYZ}^{ABC} (t) = \mathcal{U} \, ( \varrho^{ABC} \otimes \varrho^{XYZ} ) \, \mathcal{U}^\dagger \, ,
\label{Eq:TE}
\end{eqnarray}
where $\mathcal{U} = \hat{U}_k^A \otimes \hat{U}_k^B \otimes \hat{U}_k^C$ is the combined unitary operator modeling the interaction of each 
atom-field pair. The density matrix of three atoms only can be obtained by taking partial traces over cavities, that is,  
\begin{eqnarray}
\varrho^{ABC} (t) = Tr_{XYZ} \big[ \, \varrho_{XYZ}^{ABC} (t) \, \big] \,  .
\end{eqnarray}
In fact we can partial trace over any systems to obtain the density matrix for the remaining parties. The general solution is simple to obtain, however, 
the calculations are quite tedious and lengthy as it is well known for Jaynes-Cummings interactions. Due to this reason, we do not provide 
the general solution here. 

\section{Detection of separability, bi-separability, and genuine multipartite entanglement} 
\label{Sec:GME}

For bipartite systems a given quantum state can be either entangled or separable, whereas for multipartite systems it is not that much
simple. It is well known that a given multipartite state can be either fully separable, bi-separable or genuinely entangled. This last possibility is 
the peculiarity of multipartite systems. For bipartite states if the partial transpose with respect to any part of a density matrix is negative 
then the state is guaranteed to be entangled, whereas for multipartite states it is known that there are quantum states which have negative partial 
transpose with respect to all bipartitions, nevertheless are not genuinely entangled \cite{Horodecki-RMP-2009, gtreview}. 

Here, we briefly present some of the criteria to detect multipartite correlations. It is sufficient to consider three qubits $A$, $B$, and $C$, 
with obvious extension to $N$ parties and higher dimensions. A three-qubit state is called fully separable if it can be written as 
$\varrho^{fs} = \sum_i p_i \varrho^A \otimes \varrho^B \otimes \varrho^C$. A pure state is bi-separable if 
$|\psi^{bs}\rangle = |\phi^A\rangle |\phi^{BC}\rangle$, where $|\phi^{BC}\rangle$ is a possible entangled state for $BC$ and same for other bipartitions.
A mixed state is bi-separable if it can be written as $\varrho^{bs} = \sum_j p_j |\phi_j^{bs}\rangle\langle \phi_j^{bs}|$ for all three bipartitions.
Finally a state is said to be genuine entangled if it is not bi-separable \cite{Bastian-PRL106-2011}. 

The definitions are easy but to find out that a given quantum state is fully separable, bi-separable or genuine entangled is a hard task.
Some known conditions are as follows: Let $\varrho$ is a bi-separable three-qubit state 
then its matrix elements always satisfy the condition \cite{Guehne-NJP-2010} 
\begin{equation}
|\varrho_{18}| \leq \sqrt{\varrho_{22} \varrho_{77} } + \sqrt{\varrho_{33} \varrho_{66} } + \sqrt{\varrho_{44} \varrho_{55} }
\label{Cr:1}
\end{equation}
whereas violation mean three-qubit state is genuinely entangled. The other variant of the same criterion is 
$|\varrho_{27}| \leq \sqrt{\varrho_{11} \varrho_{88} } + \sqrt{\varrho_{33} \varrho_{66} } + \sqrt{\varrho_{44} \varrho_{55} }$, etc. 
Another relevant criterion for our work is the observation that for three-qubit bi-separable state, it holds that 
\begin{eqnarray}
|\varrho_{23}| + |\varrho_{25}| + |\varrho_{35}| \leq \sqrt{\varrho_{11} \varrho_{44} } +  \sqrt{\varrho_{11} \varrho_{66} } 
+ \sqrt{\varrho_{11} \varrho_{77} } \nonumber \\ 
 + \frac{1}{2} (\varrho_{22} + \varrho_{33} + \varrho_{55}) \,, \label{Cr:2}
 \end{eqnarray}
and violation means genuine entanglement. A full separability condition related with this criterion is 
\begin{eqnarray}
 |\varrho_{23}| + |\varrho_{25}| + |\varrho_{35}| \leq \sqrt{\varrho_{11} \varrho_{44} } +  \sqrt{\varrho_{11} \varrho_{66} } 
 + \sqrt{\varrho_{11} \varrho_{77} } \, .\label{Cnd}
 \end{eqnarray}
It should be mentioned here that these conditions are not necessary and sufficient. This means if a given quantum state does not violate bi-separability 
conditions, then we can not say with certainty that state is bi-separable.
 
Detection of genuine entanglement \cite{Bastian-PRL106-2011} can be based on using positive partial transpose (PPT) mixtures. A bipartite 
state is PPT if its partially transposed matrix has no negative eigenvalues. It is known that separable states are always PPT \cite{peresppt}. 
The set of separable states with respect to some partition is therefore contained in a larger set of states which has a positive partial transpose 
for that bipartition. Let $\varrho_{A|BC}^{PPT}$, $\varrho_{B|CA}^{PPT}$, and $\varrho_{C|AB}^{PPT}$ are PPT states with partitions shown. Then if 
a state can be written as
\begin{eqnarray}
\varrho^{PPTmix} = p_1 \, \varrho_{A|BC}^{PPT} + p_2 \, \varrho_{B|AC}^{PPT} + p_3 \, \varrho_{C|AB}^{PPT}\,, 
\end{eqnarray}
then it is called a PPT mixture. As any bi-separable state is a PPT mixture, therefore any state which is not 
a PPT mixture is guaranteed to be genuinely entangled. PPT mixtures can be fully characterized by semidefinite programming 
(SDP), and the set of PPT mixtures is a very good approximation to the set of bi-separable states. 
However, it should be noted that there are genuine entangled states which are PPT mixtures \cite{Bastian-PRL106-2011}. So if the below measure is 
positive for a given state then it is guaranteed to be genuine entangled, if not then we can not say with certainty whether the state is 
bi-separable or genuinely entangled. 

It was proved \cite{Bastian-PRL106-2011} that a state is a PPT mixture iff the following optimization problem 
\begin{eqnarray}
\min {\rm Tr} (\mathcal{W} \varrho)
\end{eqnarray}
with constraints that for all bipartition $M|\bar{M}$
\begin{eqnarray}
\mathcal{W} = P_M + Q_M^{T_M},
 \quad \mbox{ with }
 0 \leq P_M\,\leq 1 \mbox{ and }
 0 \leq  Q_M  \leq 1\, 
\end{eqnarray}
has a positive solution. The constraints just state that the considered operator $\mathcal{W}$ is a decomposable entanglement witness for any 
bipartition. If this minimum is negative, then $\varrho$ is not a PPT mixture and hence is genuinely multipartite entangled. The absolute value 
of the this minimization is an entanglement monotone for genuine entanglement \cite{Bastian-PRL106-2011}. We denote the measure by $E(\varrho)$. 
For bipartite systems, this monotone is equivalent to {\it negativity} \cite{Vidal-PRA65-2002}, whereas for multipartite systems, it can be called 
genuine negativity. 

\section{Results} \label{Sec:results}

In this section, we present two key observations. First, by taking correlated cavities and atoms to be in entangled state, we demonstrate that 
three atoms can achieve genuine entangled state. Second, we show explicitly that if cavities $Y$ and $Z$ are classically correlated then resulting 
state of three atoms is bi-separable. However, if cavities are quantum correlated, then we can not comment on entanglement properties of atomic state. 
However, based on relative difference between diagonal and off-diagonal matrix elements, we guess it to be bi-separable. 

\begin{figure}[h]
\scalebox{2.35}{\includegraphics[width=1.99in]{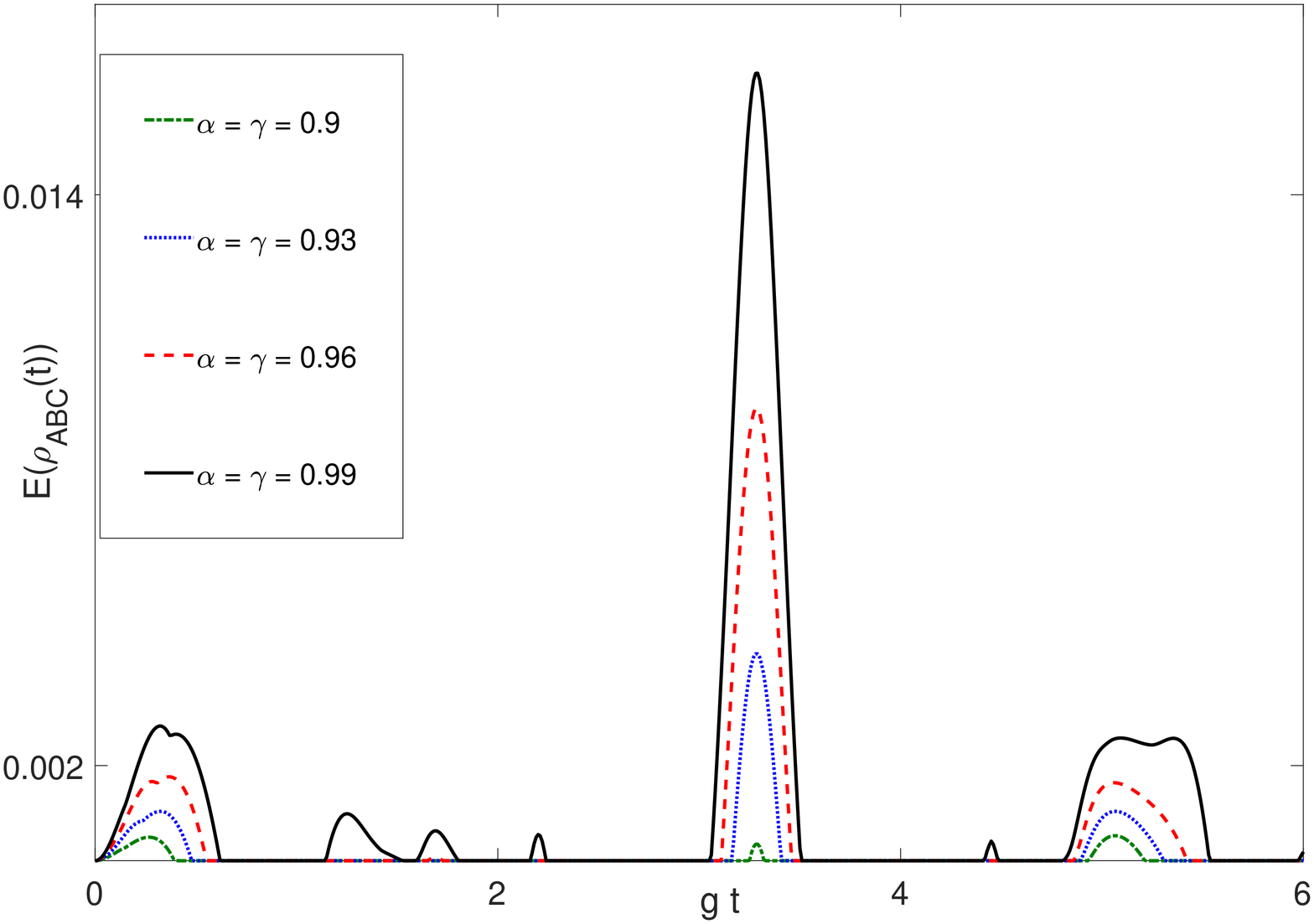}}
\caption{(Color online) genuine entanglement for three atoms is plotted against parameter $g t$ for various settings of parameters $\alpha$ and $\gamma$. 
We can see that all three atoms become genuinely entangled. See text for details.}
\label{FIG:ABCgme}
\end{figure}

Let us take atoms $A$ and $B$ (as well as cavities $Y$ and $Z$) to be entangled. Although this can be achieved by taking parameter $\alpha > 1/3$ 
($\gamma > 1/3$), however, we set them larger than $0.9$ to demonstrate the genuine entanglement. As the atom $C$ develops entanglement with cavity 
$Z$ once the interaction is on, we expect that it can also develop entanglement with other atoms via this indirect interaction. 
This is precisely what we observe in Figure~(\ref{FIG:ABCgme}), where genuine entanglement for three atoms is plotted against parameter $g t$ for 
various settings of parameters $\alpha$ and $\gamma$. We set $\beta = \kappa = 1/\sqrt{2}$ for simplicity, although we can take any other range as well. 
We conclude two main results from this figure. First, it is a clear demonstration that it is possible to genuinely entangle three uncorrelated atoms 
via JC-interactions. Second, numerical value of genuine entanglement for state $\varrho_{ABC}(t)$ depends on 
$\alpha$ and $\gamma$ as we decrease these parameters, genuine entanglement is also decreased but periodic behavior is similar. However, we want to 
clarify again that this measure of genuine entanglement is not a necessary and sufficient condition. This means wherever, the measure vanishes, 
it is not possible for us to comment whether the quantum state is bi-separable or not. 
Figure~(\ref{FIG:ABCgme}) strictly demonstrate the development of genuine entanglement among three atoms.  

\begin{figure}[h]
\scalebox{2.35}{\includegraphics[width=1.99in]{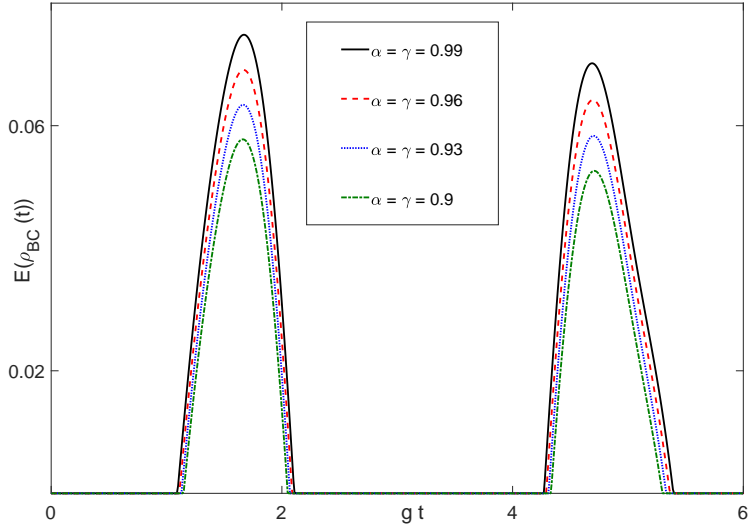}}
\caption{(Color online) Negativity is plotted for quantum state of atoms $B$ and $C$ for various settings of $\alpha$ and $\gamma$.}
\label{FIG:BCent}
\end{figure}

Now we check entanglement between atoms $B$ and $C$. We mentioned earlier that entanglement monotone in this case is equal to negativity for 
qubit-qubit systems (which is necessary and sufficient condition for entanglement). Figure~(\ref{FIG:BCent}) shows entanglement developed between 
atoms $B$ and $C$. We observe the similar periodic behavior for different settings of parameters $\alpha$ and $\gamma$, with numerical value of 
negativity decreasing with decreasing them. 

We have not detected any entanglement developed between atoms $A$ and $C$ for any range of parameters $\alpha$ and $\gamma$. Although some quantum 
correlations are developed between atoms $A$ and $C$, however, we have not computed them.

Next, we examine the initial conditions where only classical correlations are shared by atoms $A$ and $B$ (also among cavities 
$Y$ and $Z$). This can be achieved by taking parameters $\alpha = \gamma = 0$. 
In this case, we have the following observations: All of the off-diagonal elements related with GHZ-type genuine entanglement are zero, that is,
$\varrho_{18}(t) = \varrho_{27}(t) = \varrho_{36}(t) = \varrho_{45}(t) = 0$, hence no GHZ-type entanglement. For W-type entanglement, we have the 
following observation. Some of the important off-diagonal elements are also zero, like $\rho_{23}(t) = \rho_{35}(t) = \rho_{46}(t) = \rho_{67}(t) = 0$. 
The only nonzero important matrix elements are $\rho_{25}(t)$ and $\rho_{47}(t)$, however their absolute value is far low and criteria (\ref{Cr:2}) 
and (\ref{Cnd}) always hold. 
In addition, $\varrho_{ABC}(\alpha = \gamma = 0)(t)$ can be written as convex combination of bi-separable states, that is, we have been able to 
write the density matrix as
\begin{equation}
 \varrho_{ABC}(\alpha = \gamma = 0)(t) = \tilde{p}_1 (t) \, \varrho^1_{AC} \otimes |0\rangle_B\langle 0| 
 + \tilde{p}_2 (t)\, \varrho^2_{AC} \otimes |1\rangle_B\langle 1| \,,
\end{equation}
where $\tilde{p}_ 1(t) + \tilde{p}_2(t) = 1$. This clearly demonstrate that when both atoms $A$ and $B$ and cavities $Y$ and $Z$ are initially classical 
correlated then there is no way that all three atoms can develop genuine entanglement. 

We want to mention here that even classical correlations among cavities $Y$ and $Z$ are sufficient to generate some quantum correlations 
among atoms $A$ and $C$. As we find that in density matrix $\varrho_{AC}(t)$, with $\beta \neq 0, 1$, and $\kappa \neq 0, 1$, all off-diagonal elements 
are strictly non-zero. The presence of these off-diagonal matrix elements are sufficient to suggest the development of some kind of quantum correlations
among atoms $A$ and $C$. We have found that there is no entanglement generated between any pair of atoms.



Finally, we consider the initial conditions when atoms $A$ and $B$ (similarly cavities $Y$ and $Z$) are not entangled but quantum correlated, that is, 
$0 < \alpha \leq 1/3$ ($0 < \gamma \leq 1/3$). We observed that right hand sides of criteria (\ref{Cr:1}), (\ref{Cr:2}) are far greater than respective 
left hand sides. 
Genuine negativity is also zero for this range of parameters. Therefore, we are unable to comment about genuine entanglement of three atoms with 
certainty, however, we suspect that resulting states are bi-separable. In addition, we have not detected any entanglement among any pair of atoms as well.  


\section{Discussions} \label{Sec:conc}

We explored the triple JC-model for the possibility of generating multipartite entanglement among uncorrelated atoms. By considering a 
realistic situation, we examined in detail that under what conditions we can entangle three atoms non-trivially. We found that three atoms 
can be genuinely entangled if the two cavities and two atoms share entanglement. 
However, to demonstrate this possibility, we took the correlated cavities with a high fraction of pure entangled states. For lower fractions, 
it is not easy to comment on entanglement properties of resulting states.
We demonstrated explicitly that classically correlated cavities can not lead to a genuine entangled state for three atoms. 
In addition, based on density matrix for any pair of atoms, we found that even classical correlated cavities can generate quantum correlations 
among uncorrelated atoms. However, classical correlations among cavities can not generate even bipartite entanglement as we do not find any pair 
of atoms to be in an entangled state. We allowed only indirect influence via the correlations shared between two cavities.  
Finally, for the case when correlated cavities and atoms are quantum correlated, we could not figure out the entanglement properties of resulting 
state with certainty. Our results provide an option to generate genuine entangled atoms via JC-systems with no genuine entanglement at the start. 
We have present a scheme in which bipartite entanglement can be turned into multipartite entanglement. Our findings can be verified with the 
available experimental setups. 

\begin{acknowledgments}
The author is grateful to Dr. Farhan Saif for valuable discussions and Dr. S. H. Dong for sharing their work.  
\end{acknowledgments}

\end{document}